\documentclass[useAMS,usenatbib]{mn2e}

\usepackage{amstext,amsmath}

\usepackage{graphicx}	
\usepackage{amssymb}	

\def   \araa {{\rm {ARA\&A}}}
\def   \apj {{\rm {ApJ}}}

\def   \apjs {{\rm {ApJS}}}

\def   \aap {{\rm {A\&A}}}
\def   \aapr {{\rm {A\&AR}}}

\def   \mnras {{\rm {MNRAS}}}

\def   \apjl {{\rm {ApJL}}}
\def   \nat {{\rm {Nature}}}

\title[A Spiral Filament Feeding the Candidate Disc]{The ALMA view of W33A: A Spiral Filament Feeding the Candidate Disc in MM1-Main}

\author[L. T. Maud et al.]{L. T. Maud$^{1}$\thanks{E-mail: maud@strw.leidenuniv.nl (LTM)}, M. G. Hoare$^{2}$, R. Galv{\'a}n-Madrid$^{3}$, Q. Zhang$^{4}$, 
\newauthor E. Keto$^{4}$, K. G. Johnston$^{2}$, J. E. Pineda$^{5}$\\
$^{1}$Leiden Observatory, Leiden University, PO Box 9513, 2300 RA Leiden, The Netherlands\\
$^{2}$School of Physics and Astronomy, University of Leeds, Leeds, LS2 9JT, UK\\
$^{3}$Universidad Nacional Aut{\'o}noma de M{\'e}xico, Instituto de Radioastronom{\'i}a y Astrof{\'i}sica, Morelia, Michoac{\'a}n, 58089, Mexico\\
$^{4}$Harvard-Smithsonian Center for Astrophysics, 160 Garden St, Cambridge, MA 02420, USA\\
$^{5}$Max-Planck-Institut f$\ddot{u}$r extraterrestrische Physik, PO Box 1312, D-85741 Garching, Germany
}

\date{Accepted 2017 January 19. Received 2017 January 9; in original form 2016 October 20}

\pubyear{2017}

\begin{document}
\label{firstpage}
\pagerange{\pageref{firstpage}--\pageref{lastpage}}
\maketitle

\begin{abstract}

  We targeted the massive star forming region W33A using the Atacama Large Sub/Millimeter Array (ALMA) in band 6 (230 GHz) and 7 (345 GHz) to search for a sub$-$1000\,au disc around the central O-type massive young stellar object (MYSO) W33A MM1-Main. Our data achieves a resolution of $\sim$0.2$''$ ($\sim$500\,au) and resolves the central core, MM1, into multiple components and reveals complex and filamentary structures. There is strong molecular line emission covering the entire MM1 region. The kinematic signatures are inconsistent with only Keplerian rotation although we propose that the shift in the emission line centroids within $\sim$1000\,au of MM1-Main could hint at an underlying compact disc with Keplerian rotation. We cannot however rule out the possibility of an unresolved binary or multiple system. A putative smaller disc could be fed by the large scale spiral `feeding filament' we detect in both gas and dust emission. We also discuss the nature of the now-resolved continuum sources.
\end{abstract}




\section{Introduction}
There are a growing number of ALMA observations in the millimetre regime to support the existence of discs around OB-type protostars \citep[e.g.][]{SanchezMonge2014,Johnston2015}.
These agree with a scaled up paradigm of low-mass star formation beyond $\sim$10-15\,M$_{\odot}$. Generally, high-resolution and high-sensitivity observations at such wavelengths have been lacking as previous interferometric arrays could not achieve either; or have only achieved enough resolution for the most nearby sources (e.g. \citealt{Beuther2013,Maud2013b,Hunter2014,CarrascoGonzalez2012}). To date, the most convincing observations of an O-type protostellar disc are those of AFGL 4176 \citep{Johnston2015} where a clear Keplerian signature is found in the CH$_3$CN molecular line emission on $\sim$1000\,au scales around the $\sim$25\,M$_{\odot}$ source. Recently \citet{Ilee2016} also report Keplerian like rotation around a putative OB-type protostar, although it is marginally resolved. Typically Keplerian signatures are found in less massive B-type protostars \citep{Cesaroni2014,Beltran2014,SanchezMonge2014}.

In the context of star formation, discs are a key requirement to allow such massive sources to accrete
 material in the face of strong radiation pressures so that they exceed $>$10$-$15\,M$_{\odot}$ \citep{Krumholz2007,Kuiper2010,Kuiper2011}. Therefore, naively, discs should be
 seen towards many more sources than have been found. Studies with `scaled-up' models of low mass star formation applied to higher masses suggest that discs are required \citep[e.g.][]{Maud2013a,Keto2010,Johnston2011}. Observations at other wavelengths also provide evidence for discs, typically through the analysis of emission line profiles in the IR modelled as arising from the hot inner regions of Keplarian discs \citep[e.g.][]{Ilee2013}. Some multi-baseline IR interferometric studies have also suggested small $<$100\,au hot discs \citep{Kraus2010,Boley2013}.

The recent review by \citet{Beltran2016} provides a comprehensive
summary of discs around luminous YSOs. In particular, they note that B-type protostars appear to have traits of scaled-up low-mass protostars, whereas, the most massive early-O type protostars (L$>$10$^{5}$\,L$_{\odot}$) are found to have toroidal structures that may \emph{never} become stable Keplerian discs.
Cesaroni et al. (2016 - in prep.) used ALMA ($\sim$0.2$''$ resolution) to spatially resolved six of these most luminous sources. However they find a very heterogeneous sample with little evidence of Keplerian discs.

Here we detail our ALMA observations zooming into W33A (G12.91-0.26), a relatively nearby archetypal massive star formation site (2.4\,kpc, \citealt{Immer2013}). We focus on one of two previously detected strong continuum emission regions, MM1 \citep{GalvanMadrid2010} in an effort to search for a Keplerian disc around the dominating high-mass protostar designated MM1-Main (L$\sim$3.2$\times$10$^{4}$\,L$_{\odot}$\footnote{http://rms.leeds.ac.uk/cgi-bin/public/RMS\_DATABASE.cgi}). A plethora of high-resolution, sub-arcsecond, multi-wavelength observations support such a scenario. Our previous SMA data indicated a change in position of the blue- and red-shifted emission close to the peak of the continuum emission, hinting at rotation, although it is marginally resolved \citep{GalvanMadrid2010}. Other work includes: VLTI MIDI observations \citep{dewit2010} where IR emission is attributed to the hot dust from the outflow cavity walls; VLT CRIRES data \citep{Ilee2013} which attribute the 2.3\,$\mu$m CO bandhead emission to a close 1-2\,au disc; and NIR adaptive optics integral field spectroscopy \citep{Davies2010} in which a 300\,km\,s$^{-1}$ Br$\gamma$ jet along is found $\sim$perpendicular to an assumed disc seen in CO emission and absorption. 

\section{Observations}
\label{obs}
Our ALMA Cycle 1 band 6 and 7 observations (2012.1.00784.S $-$ PI: M. G. Hoare) were observed in Cycle 2 taken 
on June 24, 25 and 26, 2015. The 
PWV was $<$0.7\,mm and there was good phase stability.
Both bands had 4 spectral windows, each with a 937.5\,MHz bandwidth and
488\,kHz frequency resolution ($\sim$0.65\,km\,s$^{-1}$ at band 6
and $\sim$0.42\,km\,s$^{-1}$ at band 7).
The spectral settings covered the CH$_3$CN (J=12-11 and J=13-12)
K-ladders from K=0 to 9 in band 6, CH$_3$CN (J=19-18) K=0 to 8 and SiO (8-7) in
band 7, and had enough `line-free' channels to determine the continuum emission. The array configurations had baseline
lengths from 30.8\,m to 1300\,m in band 6 and out to 1600\,m in band 7. A minimum of 37 antennas were used. The resulting images have a beam size
of 0.32$''\times$0.26$''$ in band 6 and 0.21$''\times$0.14$''$ at band 7 using
a robust=0.5 weighting ($\sim$700\,au and $\sim$500\,au respectively). The data were manually reduced using {\sc casa} \citep{McMullin2007} version 4.5.1. 
Self calibration was performed, although a minimal
timescale of $\sim$30\,s was used as the signal-to-noise was no longer sufficient to
improve the phase residuals on shorter times.

\begin{table*}
\begin{center}
  \caption{Source parameters established from the continuum maps at the resolutions noted in Section \ref{obs} prior to free-free correction. We use apertures measuring down to the 5$\sigma$ level ($\sigma$ = 0.107 and 0.221\,mJy\,beam$^{-1}$ for band 6 and band 7) while dividing the sources at the intersection between them. The flux was summed in an aperture at the 15\,percent level for MM1-Main that was `circular' ($\sim$0.42$''$ diameter) before the emission blends into the other sources. MM1-S and MM1-NW are not resolved at 1.3\,mm. Uncertainties in the peak value can be considered as the $\sigma$ level and of the order $\sim$5$-$10\,percent for the integrated fluxes and masses. The mass is calculated using the band 7 data and shows a range for some sources \emph{after} removal of the extremes of the free-free emission. Note the total is larger than the sum of the sources as it also includes diffuse emission. The range of free-free fluxes for Main and SE are shown.}
  \vspace{-0.5cm}
{\footnotesize
\begin{tabular}{@{}lllrrrrrrr@{}}
\hline
Source  & RA.   & DEC.   & 0.87\,mm   & 0.87\,mm Peak & 0.87\,mm & 1.3\,mm  & 1.3\,mm Peak     & 1.3\,mm  & Mass    \\
 Name            & (J2000) & (J2000) &  Flux  (mJy)         & (mJy/bm) & ff (mJy)    &  Flux (mJy)     & (mJy/bm) & ff (mJy)  &   (M$_{\odot}$) \\
\hline
  Main  & 18:14:39.512 & $-$17:52:00.13 &  126.1 & 100.3 & 23.7$-$45.1 &  73.4 &  56.8  & 18.5$-$28.9  &  0.15$-$0.21\\
  S     & 18:14:39.511 & $-$17:52:00.40 &  40.2  & 16.1  & ... &  ...  &    ... & ...  &   0.10  \\
  SE    & 18:14:39.552 & $-$17:52:00.48 &  23.9   &  7.7 & 8.2$-$15.9  &  10.1 &   5.6  & 6.5$-$10.0  &  0.00$-$0.07  \\
  E     & 18:14:39.566 & $-$17:52:00.01 &  49.0  &  8.3  & ... &  18.1 &   6.1  &  ... &   0.71     \\
  NW    & 18:14:39.491 & $-$17:51:59.87 &  21.1   &  6.9 & ... &  ...  &    ... &  ... &   0.32 \\
  Ridge & 18:14:39.466 & $-$17:51:59.47 &  59.1  &  9.4  & ... &  18.2 &   5.0  & ...  &  0.58      \\
  \hline
  Total &      ...     &     ...        &  410.7   &  100.3 &  31.9$-$61.0     &   170.3     & 56.8 &  25.0$-$38.9   &   2.22$-$2.45    \\
  \hline
 \end{tabular}
}
\label{tab:tab1}
\end{center}
\end{table*}

\section{Results and Discussion}
Here we focus on the continuum and CH$_3$CN emission from W33A MM1
which harbours the high-mass protostar, as well as briefly discussing the SiO emission.
The W33A region is chemically rich with over 20 other species detected
\citep[c.f.][]{GalvanMadrid2010}, although the full discussion and modelling is beyond the scope of this work and will be presented in a forthcoming article.

\subsection{The Continuum Emission}
\label{cont_em}
Figure \ref{fig:fig1}a shows the band 7 continuum emission from the W33A MM1 region with a power-law scaling. With these high-resolution and high-sensitivity ALMA observations we further resolve the region compared to our previous work \citep{GalvanMadrid2010}. Specifically, the emission around the central brightest source MM1-Main is well resolved (peak flux of 0.96\,mJy\,beam$^{-1}$ at Band 7 without free-free correction) as the NW and SE components are now separated from MM1-Main. We also define the extra regions MM1-E and MM1-Ridge (only resolved at Band 7). 

Masses are estimated from the Band 7 data, assuming optically thin emission, using the calculated dust opacity index in the region ($\beta$ - see below) and the temperatures as discussed in Section \ref{ch3cn}. Where CH$_3$CN is too weak for our temperature estimation technique we fix it to 80\,K (the lowest temperature in MM1). We use Equation 1 of \citet{Maud2013a}, a dust to gas ratio of 100 and a dust opacity $\kappa_\nu$ = $\kappa_0 (\nu/\nu_0)^\beta$ where $\kappa_0$ = 0.5\,cm$^{2}\,$g$^{-1}$ at 230\,GHz \citep{GalvanMadrid2010}. Table \ref{tab:tab1} lists details the positions, fluxes and masses of the various sources. The masses are consistent with our SMA data \citep{GalvanMadrid2010} considering we recover nearly all the flux at the higher resolution but accounting for a higher temperature and closer source distance.

We calculate $\beta$ using the band 6 and band 7 emission with \emph{matched u,v coverage} (40 to 610 \,K\,lambda) and beam sizes \citep[c.f.][]{GalvanMadrid2010,Maud2013a} after the free-free emission is removed under the Rayleigh-Jeans approximation ($h\nu \ll k_BT$,) where the dust opacity index $\beta$ is related to the spectral index by $\alpha = \beta + 2$. We model the free-free emission by point sources for MM1-Main and MM1-SE, after extrapolating from the radio detections at 8.4, 15.0 and 43.3\,GHz \citep{vandertak2005,Rengarajan1996} with a spectral index of the free-free emission, $\alpha_{\rm ff}$=1.03$\pm$0.08 \citep{GalvanMadrid2010}. The corresponding free-free fluxes at band 6 range from 25 to 39\,mJy and at band 7 from 37 to 61\,mJy when considering $\Delta\alpha_{\rm ff}$. \citet{vandertak2005} resolve MM1-Main (Q1) and SE (Q2) and so we divide the free-free flux between the two sources according to the their 43\,GHz flux ratio.

The spectral index values $\alpha$ for MM1-Main range from 2.0 to 2.5 and the dust opacity index $\beta$ from 0.0 to 0.5 depending on the free-free contribution. $\alpha$ of $\sim$2 could point to the continuum emission becoming optically thick ($\tau>$1) within the central $\sim$500\,au. This is discrepant with the low peak brightness temperature of MM1-Main ($T_{\rm mb}\sim$27\,K) within the 0.21$''\times$0.14$''$ beam. The column density at the peak of MM1-Main reaches $\sim10^{24}\,$cm$^{-2}$, although the calculated maximal optical depth only reaches $\tau \sim$0.06 (following equations 2 and 3 from \citealt{Schuller2009}). Assuming a coupling of the gas and dust temperatures, we estimate that only the very central $\sim$30\,mas (milli-arcseconds) diameter would be optically thick, hence we are dominated by optically thin emission. Accounting for the uncertainties in the free-free spectral index and consistency with optically thin emission we emphasise that an estimated $\beta$ value of $\sim$0.5 provides good evidence for larger dust grains as a consequence of grain growth \citep{Draine2006,Williams2011}. Future modelling will bring both the density and temperature distribution to bear on the dust opacity index calculation. 

\subsection{CH$_3$CN Emission}
\label{ch3cn}
\begin{figure*}
\begin{center}
\includegraphics[width=0.82\textwidth]{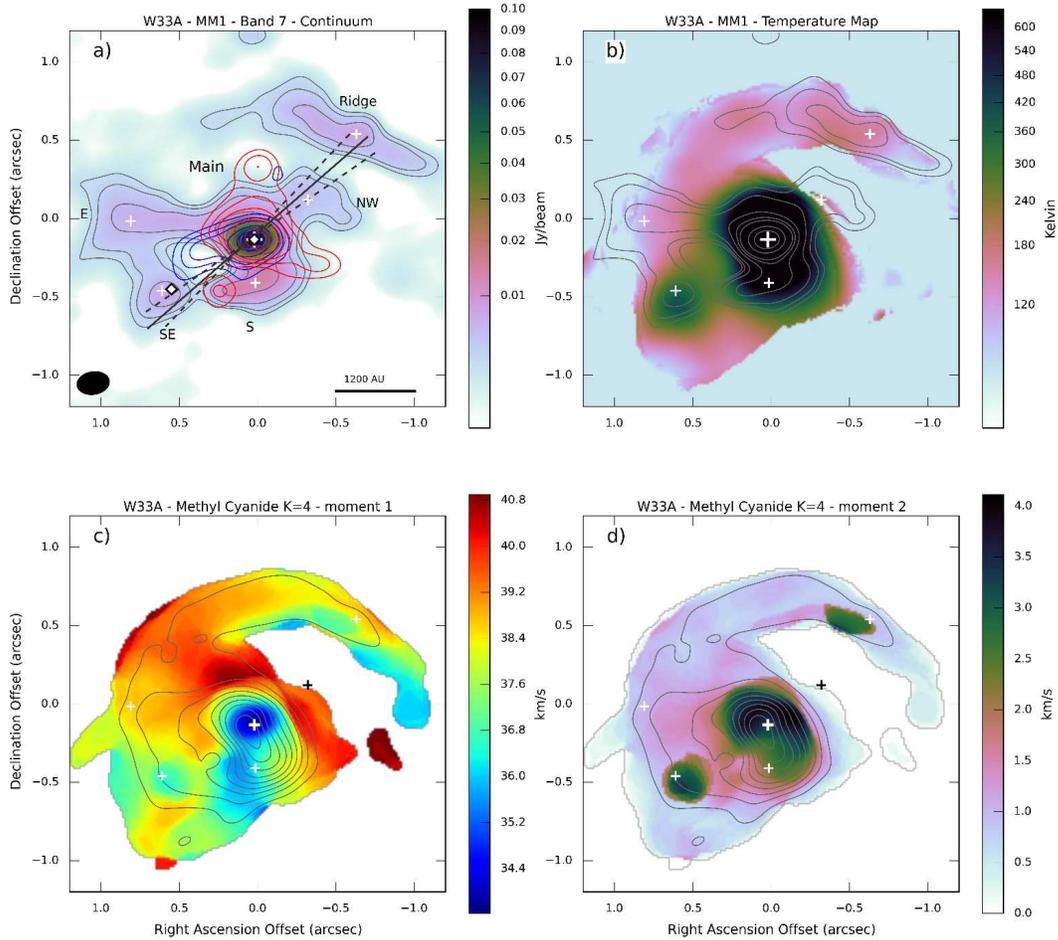}
\caption{a) 0.87\,mm continuum image of the W33A MM1 region. The mm sources are marked by crosses ($+$), the radio sources are indicated by diamonds and the CO outflow direction by the solid line with a $\pm$ 5$^{\circ}$ range (dashed lines). The grey contours are at 15, 25 50, 75 and 100 $\sigma$ ($\sigma$=0.22\,mJy\,beam$^{-1}$). The sources are labelled and a scale bar and beam are shown at the bottom of the image. The blue- and red-shifted SiO emission integrated from 29 to 36.5 and 39.5 to 43\,km\,s$^{-1}$ respectively, are represented by blue and red contours at 20 to 90\,percent in 10\,percent steps. b) Temperature map established from the fitting of CH$_3$CN. The contours are the band 7 continuum as (a). c) Moment 1 map of CH$_3$CN J=19-18 K=4 integrating velocities between 28 and 42\,km\,s$^{-1}$ above 15$\sigma$ ($\sigma$= 2.6\,mJy\,beam$^{-1}$\,per 0.42\,km\,s$^{-1}$ channel). The range of the colour bar is limited to display from 35\,km\,s$^{-1}$ as only bluer emission is located close to the peak of MM1-Main. The grey contours represent the moment 0 map of CH$_3$CN J=19-18 K=4 from 10 to 90 percent in 10 percent steps. d) The moment 2 velocity dispersion map of CH$_3$CN J=19-18 K=4. Note the spiral structure in both figures (b), (c) and (d)}
\label{fig:fig1} 
\end{center}
\end{figure*}

Figure \ref{fig:fig1}b presents the temperature map of the MM1 region using the higher spatial and spectral resolution, and signal-to-noise of the band 7 CH$_3$CN J=19$-$18 K-ladder. We follow the rotation temperature diagram (RTD) analysis outlined in \citet{Araya2005} on a pixel-by-pixel basis for a single optically thin component in LTE (see \citealt{Hollis1982,Loren1984,Turner1991,Zhang1998} for discussion). This breaks down for MM1-Main in the central region as the column density divided by the statistical weight versus the energy level no longer follows a linear trend, and so we cap the temperature at 624\,K conforming to the upper energy level of the J=19$-$18 K=8 line. With {\sc radex} \citep{vandertak2007} we can reproduce individual K-ladder line fluxes with temperatures $>$600\,K, column densities $>$10$^{17-18}$\,cm$^{-2}$ and densities $>$10$^7$\,cm$^{-3}$, but for a fixed set of these parameters we cannot reproduce the observed ratios between them. This suggests at least 2 temperature components may be required. A full radiative transfer model is required but is beyond the scope of this letter and will be presented in the future.
  
  Figure \ref{fig:fig1}c shows the moment 1 (intensity weighted velocity) and \ref{fig:fig1}d the moment 2 (velocity dispersion) maps focussing on MM1 over the full velocity range of the CH$_3$CN 19$-$18 emission from $\sim$28\,km\,s$^{-1}$ to $\sim$42\,km\,s$^{-1}$ of the K=4 line. This is representative of the K=3, 5 \& 6 line emission, but not K$>$7 which traces hotter gas closer to the central source in addition to decreasing in flux. Compared to our previous SMA observations we resolve the gas and find the kinematics are much more complex than the lower resolution observations might have indicated. We do not find a Keplerian disc \citep[c.f.][]{Johnston2015} in MM1-Main as suggested by the position shift of the blue- and red-shifted emission in \citealt{GalvanMadrid2010}, Fig. 6.

\subsection{SiO Outflow}
\label{sio}
  Strong SiO emission is detected towards MM1-Main. Clipping at $\sim$3$\sigma$ ($\sigma\sim$1.71\,mJy\,beam$^{-1}$ per 0.42\,km\,s$^{-1}$) we establish the velocity ranges from $\sim$29 to 36.7\,km\,s$^{-1}$ in the blue shifted and $\sim$39.6 to 42.8\,km\,s$^{-1}$ in the red shifted regimes (limited at blue velocities by line contamination). The blue shifted emission highlights a linear jet-like structure although the red-shifted emission is more complex in shape. The PA of the central flow based on the linear blue-shifted emission is $\sim$120$^{\circ}$ and is coincident with that of the larger scale CO outflow \citep{Maud2015,GalvanMadrid2010} and the near-IR reflection nebula \citep{Davies2010}.

\subsection{Nature of the other continuum sources}
\label{nature}
  
  It is unclear whether the components S, NW, E, SE and ridge, are embedded sources, filaments feeding the main core \citep[e.g.][]{Peretto2013}, or over-densities at the outflow cavity wall \citep[e.g.][]{Fuente2009}. Considering the direction of the jet/outflow (SE$-$NW) then one plausible description is that the extensions E and S (curved line through Main) could form part of a blue-shifted cavity wall, whereas NW would form part of the red-shifted cavity wall. Certainly the moment 1 map Figure \ref{fig:fig1}c indicates the blue and redshifted emission is not confined only to the NE and SW but is co-located with the outflow direction SE-NW \citep{Maud2015,dewit2010}. The moment 2 map also highlights the larger velocity dispersions in the region of the blue-shifted outflow direction tracing the cavity wall working surfaces. Such a description corroborates the faint $<$10\,percent level dust emission in the region of MM1-E and -S as from an outflow cavity. However Figure \ref{fig:fig1}b indicates that S has additional dust emission and elevated temperatures consistent with central heating and also has broader emission lines comparable to those in MM1-Main. A highly probably alternative scenario is that S is an embedded intermediate-high mass source providing the heating and velocity dispersion although it is not separated in the continuum emission.

  The mm emission from SE appears in the path of the SiO jet and can attributed to that from a thermal jet `knot', as the mm emission is accounted for by the extrapolated free-free emission. Such an interaction is supported by the elevated temperatures at the `knot', although the line profile do not show outflow like wings and there is no SiO contribution at this location. Considering the lower limit of the free-free contribution to SE at band 7 leaves some mm emission as from dust and assuming the free-free emission as a tracer of star-formation activity itself, we believe SE could also be explained by an embedded self heating source. SE is very rich in complex chemical species which may be more closely related to protostellar heating although these tracers could still be generated by grain sputtering due to outflow activity \citep[e.g.][]{Arce2007,Arce2008,Codella2009,Drozdovskaya2015}.
  
  MM1-Ridge could potentially harbour a protostellar source, or simply be an enhancement associated with the red-shifted outflow lobe. However, the whole structure has a clear velocity gradient (Figure \ref{fig:fig1}c) from a blue-shifted velocity of $\sim$37\,km\,s$^{-1}$ at the tail of the spiral shape to the south of ridge, near $V_{\rm LSR}$ velocities flowing through ridge, and finally onto the north of MM1-Main at red-shifted velocities $\sim$41\,km\,s$^{-1}$. Based on these kinematics and the spiral structure we strongly suggest that it is a `feeding' filament drawing material from the reservoir to the South-West (joined to MM2 NE and MM2 $-$ \citealt{GalvanMadrid2010} and in our continuum emission at the 5$-$10\,percent flux levels). Interestingly similar spiral-like structures were seen in the simulated images of \citet{Krumholz2007} who modelled the formation of a massive stars via discs. The spirals in their work feed and join the $<$500$-$1000\,au size discs surrounding the compact, multiple cores formed in the dynamic environment. Observations of the low-mass binary L1551 and the high-mass proto-cluster G33.92 indicate such accretion/infall spiral features maybe commonplace \citep{Takakuwa2014,Liu2015}.

\subsection{W33A MM1-Main as a disc candidate?}
\label{idea}
We do not find a Keplerian rotation signature at $\sim$500\,au resolution. The spectra of all the unblended lines of the CH$_3$CN ladders in the region of MM1-Main are not Gaussian but are very broad. The profile averaged over the central 1000\,au appears double peaked, indicative of a velocity shift with position, hinting at an underlying rotation. Under a naive assumption that the velocity shift is due to Keplerian rotation we establish a rough stellar mass estimate after fitting the CH$_3$CN lines at the flux peaks of the blue- and red-shifted emission separated by $\sim$1000\,au in the NE-SW direction (centroids at $\sim$34.8 and $\sim$42.5\,km\,s$^{-1}$). The central source would be at least 13\,M$_{\odot}$ accounting for the inclination angle of $\sim$50$^{\circ}$ \citep{dewit2010}. We cannot exclude that this emission is from two separate components, possibly an unresolved binary system. The moment 2 map (Figure \ref{fig:fig1}d) indicates the inner 500-600\,au region as a hive of activity where the velocity dispersion of exceeds 4\,km\,s$^{-1}$ and could point to the a disc on smaller scales that joins with the spiral `feeding filament' we see on the scales observed.

We find the MM1 region is complex; chemically, kinematically and morphologically. The enhanced cavity emission, other embedded sources and spiral `feeding filament' have unique kinematic signatures that could either mask a Keplerian disc signature, or themselves inhibit the formation of a disc. The spiral structures in our observations appear very similar to those seen in the simulations of \citet{Krumholz2007}, which show the formation of a multiple-protostellar system via discs and feeding spirals. For W33A MM1-Main the disc would be less than 500\,au in size as constrained by our observations.

\section{Summary}
  We present the highest resolution observations to date ($<$0.2$''$, $\sim$500\,au) of W33A MM1 at sub-mm wavelengths. These resolve the emission surrounding the MM1 core into components associated with cavity wall
  emission, other possible protostars and a spiral shaped feeding filament. The CH$_3$CN emission used to calculate the temperatures in the region
  indicate it is in excess of 600\,K for MM1-Main. Although there is some evidence of some rotational motions around the central O-type protostar, W33A MM1-Main, there is no simple Keplerian accretion disc on scales below 1000\,au as have been detected in similar sources. We find a spiral structure that, when considering models and other data, suggests a more dynamic environment where material could feed to a smaller scale $<$500\,au disc. The entire star forming region warrants a higher resolution study taking advantage of the ALMA $\sim$15\,km baselines. 

  \vspace{-0.3cm}
\section*{Acknowledgements}
This paper makes use of the following ALMA data: ADS/JAO.ALMA\#2012.1.00784.S. ALMA is a partnership of ESO (representing its member states), NSF (USA) and NINS (Japan), together with NRC (Canada), NSC and ASIAA (Taiwan), and KASI (Republic of Korea), in cooperation with the Republic of Chile. The Joint ALMA Observatory is operated by ESO, AUI/NRAO and NAOJ. R.G.-M. acknowledges support from UNAM-DGAPA-PAPIIT IA101715. JEP acknowledge the financial support of the European Research Council (ERC; project PALs 320620). 




\bibliographystyle{mn2e}





\end{document}